# Broadband near-ultraviolet dual comb spectroscopy

Lukas Fürst (1), Adrian Kirchner (1), Alexander Eber (1), Florian Siegrist (1), Robert di Vora (1), Birgitta Bernhardt (1)*

*Correspondence to: bernhardt@tugraz.at

(1) Institute of Experimental Physics, Graz University of Technology, Petersgasse 16, 8010 Graz, Austria

## Abstract

The highly energetic photons of ultraviolet light drive electronic and rovibronic transitions in all molecular species. This radiation is thus a prime tool for strongly selective spectroscopic fingerprinting and real-time environmental monitoring if broad spectral coverage, short acquisition times and high spectral resolution is achieved – requirements that are in mutual competition in traditional applications. As a novel approach with intrinsic potency in all three aspects, here we introduce ultraviolet dual comb spectroscopy using two broadband ultraviolet frequency combs centered at 871 THz and covering a spectral bandwidth of 35.7 THz. Within a 100 µs acquisition time window, we obtain rotational state-resolved absorption spectra of formaldehyde, a prototype molecule with high relevance for laser spectroscopy and environmental sciences. This is the first realization of broadband dual comb spectroscopy in the ultraviolet spectral region and a pioneering tool to allow for real-time monitoring of rovibronic transitions.

## 1. Introduction

A vast variety of photochemical reactions are induced by ultraviolet (UV) radiation. Specifically, solar UV radiation triggers many atmospheric reaction and fragmentation processes that involve environmental trace gas species like $O_3$, $NO_2$, NO and HCHO. In order to improve our understanding of the relevant reaction pathways, spectroscopic access to the UV spectral region is of paramount importance. Because most molecular gases show strong and congested absorption characteristics in the ultraviolet region [1,2], a high spectral resolution is required for complete state determination. At the same time, broad spectral coverage is desired to achieve high specificity and to simultaneously detect characteristic transitions of reactants and products of a photochemical reaction. Finally, rapid acquisition times can grant access to the exploration of reaction rates. So far, the combination of all three criteria - high spectral resolution (50 GHz), broad spectral coverage (>30 THz) and short acquisition times (<10 s) - has not yet been accomplished in the UV region. State-of-the-art techniques aiming at high spectral resolution involve scanning techniques, providing instrumental linewidths on the order of 100 MHz, i.e. resolving powers up to $10^7$, with measurement times of minutes to hours [3–8]. Alternatively, experimental observations aiming at a broad bandwidth and high spectral resolution have been shown to cover up to 533 THz within 120 s measurement time and exhibit a resolving power of $9.6\times10^3$ at a spectral resolution of 83 GHz [1,9,10].

In this work, we introduce ultraviolet dual comb spectroscopy (DCS) that simultaneously features fast acquisition, broad detection bandwidth and high resolution as a novel tool suitable for studying complex photo-chemical reaction processes in gaseous media. DCS is an innovative spectroscopic method [11–14] that has proven its capabilities in numerous implementations operating in the visible spectral region across the infrared down to the THz domain [12,15–20]. Especially, mid infrared DCS can be exploited for sensitive and accurate determination of rotational and vibrational transitions, the so-called molecular fingerprints [21–23].

In contrast, the UV region has so far been relatively unexplored by DCS, mainly due to the lack of readily available laser frequency comb sources [24–26]. Most recently, several groups are working on establishing DCS in the UV spectral region [27–30]. This is because introducing DCS with its rovibronic detection capabilities into the UV region, implying high photon energies, would additionally enable fingerprinting the electronic energy structure of matter and thus help to unravel a fundamental riddle in spectroscopy: the complex trinity of rotational, vibrational and electronic excitation in gaseous media. Excitations involving electrons can be followed by various reaction pathways and their resulting different molecular components are highly relevant, like the $NO_x$ cycle for atmospheric sciences [5].

In this work, we perform fast and broadband UV DCS of formaldehyde (HCHO). HCHO plays a primary role in tropospheric chemistry and is the most abundant and most important organic carbonyl compound in the earth`s atmosphere [31]. Due to its high environmental relevance and multifaceted absorption characteristics, HCHO has been studied spectroscopically for decades under different experimental conditions aiming either at a high spectral resolution, broad spectral coverage or high sensitivity [3–5,9,32–34]. We demonstrate the coalition of those important spectroscopic criteria with single-trace DCS in the UV, i.e. one interferogram only. Averaging multiple DCS traces enables a higher signal-to-noise-ratio (SNR). As a result of averaging, longer apodization windows are possible, achieving an improved spectral resolution but with the same SNR when compared to the single-trace spectra. Finally, we exploit the strong UV absorption cross section enabling single-trace acquisition for real-time UV DCS realizing ad-hoc monitoring of the HCHO concentration in our sample cell.

## 2. Methods

### 2.1 Experimental scheme

Figure 1a illustrates the experimental setup, employing two Ytterbium-based frequency comb oscillators. Their outputs are centered at a wavelength of 1030 nm, and their repetition rates are stabilized to $f_{rep,I}$ = 80 MHz and $f_{rep,II}$ = 80 MHz + 6 Hz, respectively. After amplifying the average power in a pulse-picking fiber amplifier and reducing the repetition rates by a factor of eight, we generate the third harmonic of the amplified frequency combs by a collinear, two-step nonlinear upconversion process (see Appendix). Figure 1b shows the resulting Gaussian-shaped UV spectra centered at $f_c$ = 871.2 THz (344 nm wavelength) with a full width at half maximum (FWHM) of 2.6 THz and 3.1 THz, respectively. Then, we inject the UV beams into a 10 cm long solid-core silica fiber for spectral broadening.

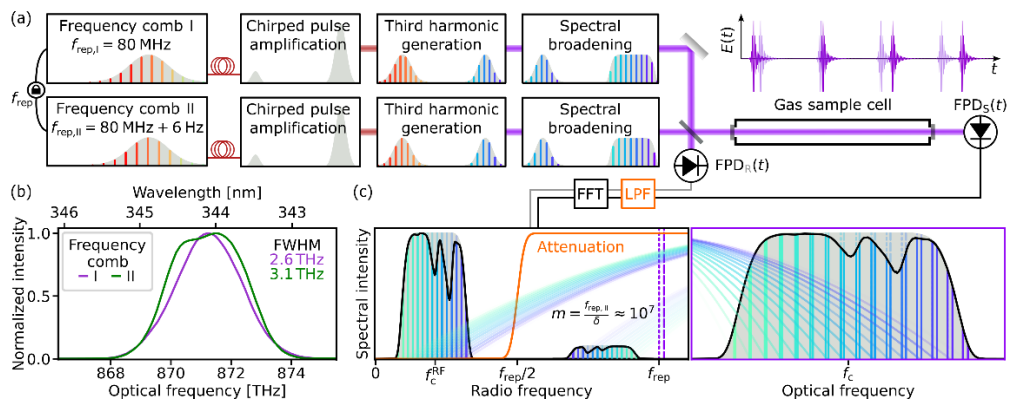

Fig. 1. Experimental implementation of dual comb spectroscopy in the ultraviolet spectral region. (a) The outputs of two laser frequency combs (commercial Yb fiber laser oscillators), stabilized in their repetition frequencies $f_{rep}$ are seeding two pulse-picking Yb fiber amplifiers, reducing the repetition rates of the oscillator pulse trains by a factor of eight and supplying Watt-level average powers at ~10 MHz repetition rates. These two high-power near-infrared laser frequency combs are frequency upconverted via third harmonic generation in nonlinear crystals. Subsequently, the UV light is spectrally broadened in a solid core silica fiber. After spatial superposition, fast photodetectors (FPD) record the time-domain interferograms of the sample (S) and the reference (R) path, respectively. (b) Measured frequency upconverted laser spectra centered at $f_c$ = 871.2 THz with a FWHM of 2.6 THz and 3.1 THz, respectively. (c) Schematic representation of a dual-comb spectrum after fast Fourier transformation (FFT). Computing the FFT of the low-pass-filtered (LPF) interferograms yields the downconverted optical transmission spectrum in the radio frequency (RF) domain. We calculate the downconversion factor m by dividing the repetition rate $f_{rep,II}$ by the detuning of the two frequency combs δ, which is used to convert the spectral information back to the ultraviolet region.

For dual-comb experiments, the two spectrally broadened output beams are superimposed using a 50:50 beam combiner. The two frequency combs with detuned repetition rates correspond to two pulse trains with a sweeping delay between consecutive pulses. This optical delay is analogous to the varying pathlength difference in Fourier-transform spectroscopy using a scanning-mirror interferometer, resulting in a downconverted heterodyne signal. The obtained time-domain interferogram has its largest value at zero delay between the two pulses due to maximum constructive interference. After the fast Fourier transformation (FFT) of the interferogram, we obtain a spectrum in the radio frequency domain at $f_c^{RF}$, which can be converted back to the ultraviolet region by multiplying it with the downconversion factor m = $f_{rep,II}/δ$, with δ being the repetition rate detuning (see Fig. 1c).

After the combiner, one output is sent through a multi-pass sample cell yielding an interaction path length of 2.3 m and focused onto a fast photodetector. The second beam exiting the combiner is focused directly onto a second fast photodiode for referencing. We record both dual-comb interferograms simultaneously (see Supplementary).

*2.2 Nonlinear spectral broadening in the UV region*

To extend the spectral coverage, we employ nonlinear spectral broadening via self-phase modulation in a silica fiber. Figure 2a shows the spectral evolution of the 168-fs-long UV input pulses simulated by solving the generalized nonlinear Schrödinger equation to find the optimum fiber length [35]. We determine a 10 cm-long fiber to yield maximum spectral coverage while keeping the pulse short. Using this fiber length, we characterize the gain in spectral bandwidth by measuring the output spectrum with a grating spectrometer at different input pulse energies (see Fig. 2b) and observe symmetric broadening of the input spectrum in excellent agreement with the simulation. The maximum bandwidth is achieved for an input pulse energy of 3.5 nJ. Higher output powers are advantageous as we are mainly limited by detector noise (see Supplementary). However, at higher input pulse energies non-reversible photodarkening of the fiber occurs [36]. The spectral bandwidth at the 6.5-µW/THz-level agrees with the simulation and scales logarithmically with input pulse energy (see inset Fig. 2b). We achieve a maximum bandwidth of $\Delta f$ = 50.3 THz and a spectral broadening factor of 12.5, surpassing previous results in the UV by a factor of two [37,38]. For UV DCS absorption experiments, we use an input pulse energy of 1.7 nJ, resulting in a spectral coverage of 35.7 THz which extends over three full vibronic absorption bands of formaldehyde centered at 861 THz, 874 THz and 885 THz, respectively (see Fig. 2b).

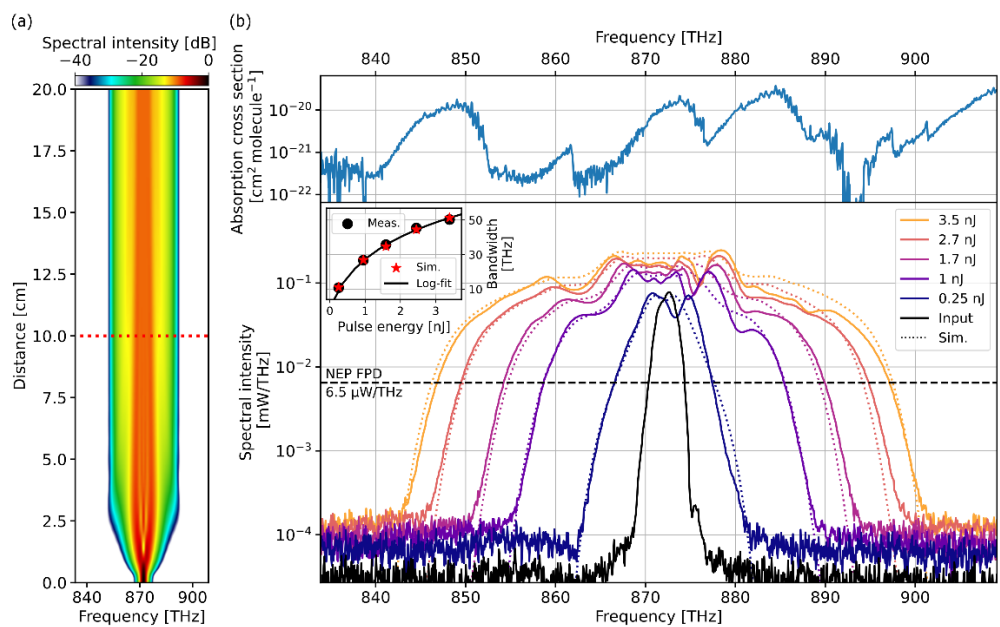

Fig. 2. Direct spectral broadening in the ultraviolet region using a silica fiber. (a) Calculated propagation distance-dependent gain in spectral coverage [35]. Red dotted line: Fiber length of 10 cm leading to maximum bandwidth and minimal deterioration of the temporal pulse shape. (b) Bottom: Output spectra of a 10 cm long fiber at different pulse energies of the UV seed laser. Dotted lines: Simulated (Sim.) output spectra using the same pulse energies as input parameters. Black solid line: Input spectrum measured directly in front of the fiber. Inset: The spectral bandwidth of measurement and simulation are in excellent agreement and scale logarithmically with the input pulse energy. The bandwidth is defined at the 6.5-µW/THz-level, which corresponds to the noise equivalent power of the fast photodetectors (NEP FPD, black dashed line, see Supplementary). The maximum spectral coverage of 50.3 THz was achieved for 3.5 nJ input pulse energy, accomplishing a broadening factor of 12.5. Top: Absorption cross section of formaldehyde [1]. The broadened spectrum fully covers the absorption bands centered at 861 THz, 874 THz and 885 THz, respectively, i.e. the $4^2_0$, the $4^3_0$ and the $2^1_0 4^1_0$ vibronic branches of the electronic transition $\tilde{A}^1 A_2 - \tilde{X}^1 A_2$.

## 3. Results

*3.1 Dual-comb interferometry in the near-UV spectral region*

The resulting interferogram is recorded by an oscilloscope and depicted in Figure 3a. Figure 3b shows a close-up view of the centerburst, which reveals oscillations arising from the interference of the two detuned pulse trains. After the FFT of the 100 µs-long interferogram, we obtain the downconverted laser spectrum in the radio frequency domain (see Fig. 3c).

We perform a sliding-window analysis of the centerburst to characterize the time dependency of our dual comb spectrum (see Fig. 3d). For each window position, the interferogram trace is apodized using a 3-µs-long super-Gaussian time window with subsequent FFT. The intense part of the spectrum up to the $1/e^2$ value is contained in a 4-µs-long time interval centered around zero delay. For window positions $\tau < 2$ µs, we observe a temporal wing, which emanates from the uncompensated chirp after spectral broadening via self-phase modulation in the fiber (see Supplementary). For DCS, we adjust the average power in front of each fiber to compensate for the different transmissions through the individual optical paths, yielding a balanced signal strength after the beam combiner. This asymmetric fiber input configuration results in a tilted spectrogram with unequal intensity distribution of the temporal wings (see Supplementary), not affecting the data analysis as the contributions are identical for both sample and reference path.

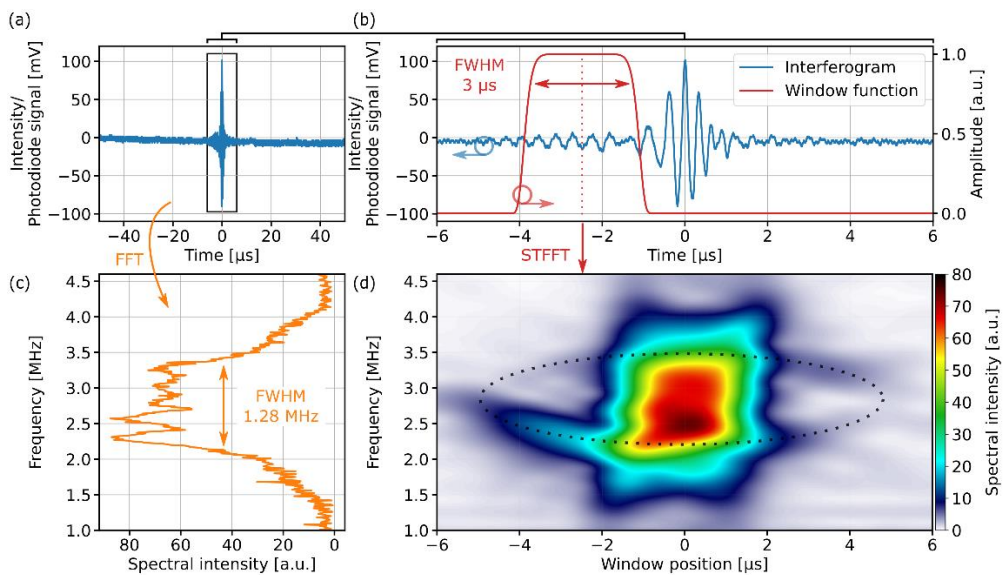

Fig. 3. Dual-comb signals generated from two interfering UV frequency combs. (a) Time-domain interferogram generated by two UV frequency combs and recorded with a fast oscilloscope. (b) Close-up view of the interferometric centerburst with characteristic oscillations arising from the interference of the detuned pulse trains. Red: 3-µs-long super-Gaussian apodization time window. (c) Downconverted laser spectrum, centered at $f_c^{RF}$ = 2.8 MHz, obtained after the fast Fourier transform (FFT) of the 100-µs-long time trace. The spectrum covers $\Delta f^{RF}$=1.28 MHz (FWHM) in the radio frequency domain. (d) Spectrogram generated by calculating the short-time fast Fourier transformation (STFFT) for different apodization time window positions displayed with linear colorscale. For window position $|\tau| > 2$ µs, temporal wings are observed, which form an elliptic pattern around the center part (indicated by black dashed line).

*3.2 Ultraviolet dual comb spectroscopy of formaldehyde*

Figure 4a compares the UV-DCS recorded absorbance spectrum of HCHO to values reported by Bass et al. [1]. The absorbance spectrum exhibits congested absorption characteristics with several vibronic excitation bands featuring rotational substructure superimposed to the electronic transition driven by the UV photons (see Fig. 2b top) [1,39]. Within only 1620 µs apodization time (corresponding to 6 single traces of 270 µs duration each, see Supplementary) we achieve 50 GHz spectral resolution, i.e. a resolving power of $1.8 \times 10^4$, across the entire 35.7 THz bandwidth covering three vibronic branches of the molecule. The measurement unveils a plethora of narrow lines arising from rotational states allowing to assign the full set of quantum numbers to the rovibronic transition as exemplified for six prominent resonances in Fig. 4a [40].

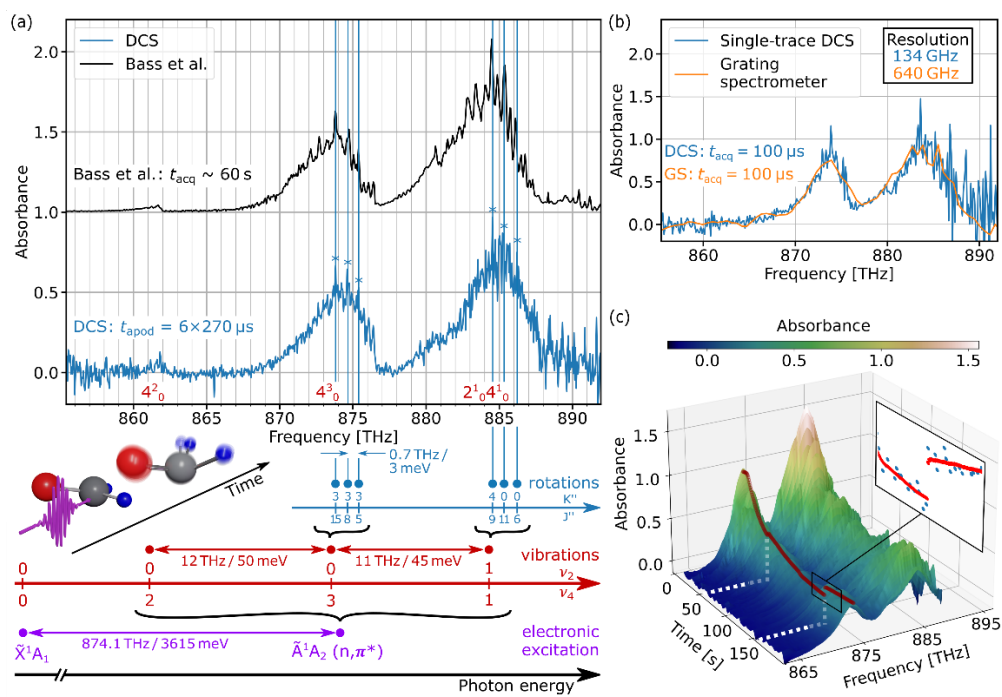

Fig. 4. Fast and broadband UV DCS resolving rotational energy levels of formaldehyde. (a) UV absorbance spectrum of formaldehyde measured by DCS (blue) and by Bass et al. (black) [1], both clearly revealing the rotational substructure of the energy levels. The dual-comb trace was obtained by averaging six traces resulting in a total apodization time of 1620 µs (total acquisition time of 8 s). The high resolution allows the assignment of all relevant quantum numbers to the absorption peaks implying characterization of the complex trinity of rovibronic energy states [40,41]. These three types of transitions induce a non-planar orientation (electronic excitation) of the formaldehyde molecule together with a superposed vibrational and rotational motion [39]. (b) Single-trace absorbance spectrum measured with DCS (blue) and a grating spectrometer (GS, orange), both achieved with the same acquisition time window of 100 µs. The DCS curve yields more than a factor of four higher spectral resolution, featuring single-trace sampling of electronic fingerprints resolving rotational states, e.g. the absorption peak at 874 THz. (c) Real-time DCS series during evacuation of the sample cell pre-filled with formaldehyde as an application of real-time gas concentration monitoring. Lineout: Expected absorbance (red dots) at 874 THz together with the value of the DCS peak absorbance (blue dots). After 50 s and 152 s the valve to the reservoir is opened intentionally to introduce abrupt changes in the sample concentration. This is observed in the dual-comb signal, demonstrating the instant response of our UV DCS monitor.

Figure 4b corroborates the resolving power of our UV-DCS concept even further. For single-trace dual comb spectroscopy the measurement time can be faster than the inverse of the repetition rate detuning (here: 1.3 seconds, see Supplementary), which defines the duration between consecutive interferograms. Here, we record one single trace in 100 µs achieving 134 GHz spectral resolution (see Supplementary). This reduced acquisition time window is sufficient to resolve the rotational absorption peaks, e.g. at 874 THz corresponding to the rotational level with K´´ = 3 and J´´ = 15 in the $4^3_0$ vibronic band of the $\tilde{A}^1 A_2 - \tilde{X}^1 A_2$ system (see Supplementary Table 1), thus demonstrating fast acquisition and high spectral resolution simultaneously. In comparison, this rotational substructure cannot be identified using a state-of-the-art grating spectrometer recording with 640 GHz resolution under identical experimental conditions (blue and orange lines, respectively).

### 3.3 Real-time monitoring of gas concentration changes

To demonstrate the potential of UV-DCS for rapid-update environmental monitoring, we monitor the pressure variation in the gas sample cell (see Fig. 4c and S5). While acquiring dual-comb interferograms consecutively as they re-occur after 1.3 s, we track the pressure of formaldehyde to compute the expected absorbance change. The absorbance at 874 THz precisely follows the calculated reference points, with a single-trace noise equivalent absorption of $NEA = 5.9 \times 10^{-4}\ cm^{-1}\ Hz^{-1}$ (pressure 0.21 mbar, see Supplementary). A standard computer derives the absorbance spectrum from the interferograms in less than 200 ms, rendering the update rate fully applicable for real-world applications, e.g. the formaldehyde emission monitoring in the wood and textile industry where, typically, HCHO concentration levels up to 4000 ppm, i.e. partial pressure of 4 mbar at atmospheric conditions, accumulate [42,43]. The temporal resolution of UV DCS can be dramatically improved to the femtosecond scale by the implementation of pump-probe schemes [44,45]. Since UV DCS is capable of rapid fingerprinting the electronic energy structure, on line monitoring of ultrafast complex reaction pathways on the femtosecond scale comes into reach.

## 4. Discussion

For quantitative comparison, we define the parameter $M_T = \Delta f/(\delta f \cdot \sqrt{T})$ with the bandwidth $\Delta f$, the spectral resolution $\delta f$ and the acquisition time T, in analogy to the work by Newbury et al. [46]. The quality factor $QF = SNR \cdot M_T$, with SNR being the signal-to-noise ratio in the frequency domain (see Supplementary), is determined to a value of $5.9 \times 10^3 \sqrt{Hz}$ for the averaged case. Thereby, we achieve a QF about three times higher than another recent UV dual-comb system employing photon counting [47]. Notably, the work by Xu et al. successfully targets at high spectral resolution while the experiments presented here focuses on a broader spectral coverage. Hence, a fair comparison remains difficult due to the different parameter space of the two demonstrations (see Supplementary).

Future developments aim at increasing the repetition rate, which will allow faster gas monitoring by using higher detuning values. Consequently, the waiting time will be reduced when using apodization windows shorter than the re-occurrence time (see Supplementary). Limitations to the repetition rate in our setup arise due to the influence on the nonlinear processes, especially the conversion efficiency of frequency upconversion and spectral broadening. A higher repetition rate results in lower pulse energies and with that in a decreased conversion efficiency. For experiments in the NUV, this can be compensated by higher average powers of the fundamental radiation. Further development of the spectral broadening scheme will be required with increased repetition rates. Another possibility to eliminate waiting time is to periodically invert the detuning frequency of the two combs. For this approach, no adjustment of the broadening scheme would be necessary [48]. In order to investigate formaldehyde at urban abundances, the sensitivity has to be increased. This can be achieved for example by combining UV DCS with enhancement cavities [7,49].

In conclusion, this work constitutes the first demonstration of DCS in the ultraviolet spectral region utilized for broadband real-time monitoring of formaldehyde. Our system features an unprecedented combination of short acquisition time windows on the μs timescale together with GHz spectral resolution over tens of THz spectral bandwidth in the near-UV and expands the application possibilities of DCS dramatically. Since all molecular species absorb strongly in the UV, our approach can be expanded to analyze other samples, including $NO_2$ and $O_3$.

We observe the trinity of electronic, vibrational and rotational transitions in the complex absorption spectrum of formaldehyde revealing rotational state resolution while covering a full spectral bandwidth of 35.7 THz. The demonstration of real-time monitoring of the HCHO concentration forms the basis to investigate reaction pathways and rates in trace gas mixtures. The chosen example of HCHO plays an important part in the interrelated chemistries of ozone and the $HO_x$ and $NO_x$ cycles and highlights that UV-DCS will enable detailed insights into the photochemistry of our troposphere [31].

## 5. Appendix

*5.1 Experimental Design*

The laser sources in this work are based on two commercially-available Ytterbium-based fiber frequency comb oscillators with 1030 nm center wavelength, with their repetition rates stabilized to $f_{rep,I}$ = 80 MHz and $f_{rep,II}$ = 80 MHz + 6 Hz, respectively. To achieve this, both cavity lengths are controlled by a piezoelectric transducer via a feedback loop. The error signals for these loops are generated by mixing the detected repetition frequency with appropriate reference signals. For one system, it is fixed to 10 MHz, while the other reference frequency can be tuned by a waveform generator, which then yields a difference in repetition rates. The output of the stabilized oscillators is injected into the chirped pulse amplification setup.

*5.2 Chirped pulse amplification and third harmonic generation*

The two pulse-picking fiber-based amplifiers include acousto-optic modulators that reduce the seed repetition rates by a factor of eight, resulting in repetition rates of $f_{rep,I}$ = 10 MHz and $f_{rep,II}$ = 10 MHz + 0.75 Hz, respectively. The repetition rate was chosen for sufficient pulse energy available for nonlinear frequency conversion and spectral broadening. Subsequently, the two-stage, chirped-pulse amplification setups enhance the average power of the pulse trains from 5.4 mW up to 19.3 W preserving the comb structure of the seeds without adding significant phase noise [26]. With a pulse duration smaller than 250 fs (FWHM $sech^2$-fit) of the amplified pulses, we can efficiently generate the third harmonic of the fundamental light. In a first step, the second harmonic is generated in a 1.5 mm-thick Beta Barium Borate crystal. Then, a calcite plate compensates for the group delay between the two pulses followed by a zero-order waveplate, which rotates both light fields to the same polarization state. Afterwards, a second Beta Barium Borate crystal (0.8 mm thickness) is used to generate the sum frequency of the co-propagating second harmonic and fundamental radiation resulting in near-UV light centered at $f_c$ = 871.2 THz optical frequency (corresponding to a central wavelength of 344 nm, see Fig. 1c). We employ dichroic mirrors to separate the UV beam from the infrared and visible light.

*5.3 Data acquisition*

We acquire the data for DCS using a fast oscilloscope with 2 GS/s sampling rate. The voltage signals of the two photodiodes (silicon fixed gain detector, 150 MHz bandwidth) are fed to the oscilloscope via BNC cables with additional analog low-pass filters (see Fig. 1a). Prior to data acquisition, we review the spatial overlap of both beams to maximize interference signal. Then, we maximize the output power after the silica fiber via the position of the incoupling lens relative to the fiber. Afterwards, both beams are sent onto the photodiodes and data acquisition is started. We record both traces simultaneously and save them on the local memory of the oscilloscope.

**Funding.** This project has received funding from the European Research Council (ERC) under the European Union's Horizon 2020 research and innovation programme (grant agreement No 947288) and from the Austrian Science Fund FWF START programme (grant agreement no. Y1254).

**Acknowledgments.** We thank Emily Hruska and Mithun Pal for valuable feedback and Thomas Jauk for his expertise in the cover image design. We also thank Roland Lammegger for his expertise concerning the formaldehyde preparation.

**Disclosures.** The authors declare no competing interests.

**Data availability.** Data underlying the results presented in this paper are not publicly available at this time but may be obtained from the authors upon reasonable request.

**Code availability.** The code used to generate the figures is available from the corresponding author upon request.

# Supplementary Materials for

## Broadband near-ultraviolet dual comb spectroscopy

Lukas Fürst *et al.*

**Interferogram simulation**

Simulations have been performed to analyze the temporal dependency of the dual-comb spectrum theoretically and compare it to the measured trace (see Fig. 3). Figure S1 depicts the spectrogram generated using unchirped pulses with balanced pulse trains as an input. The spectrogram features high symmetry with a sech$^2$-shaped spectrum.

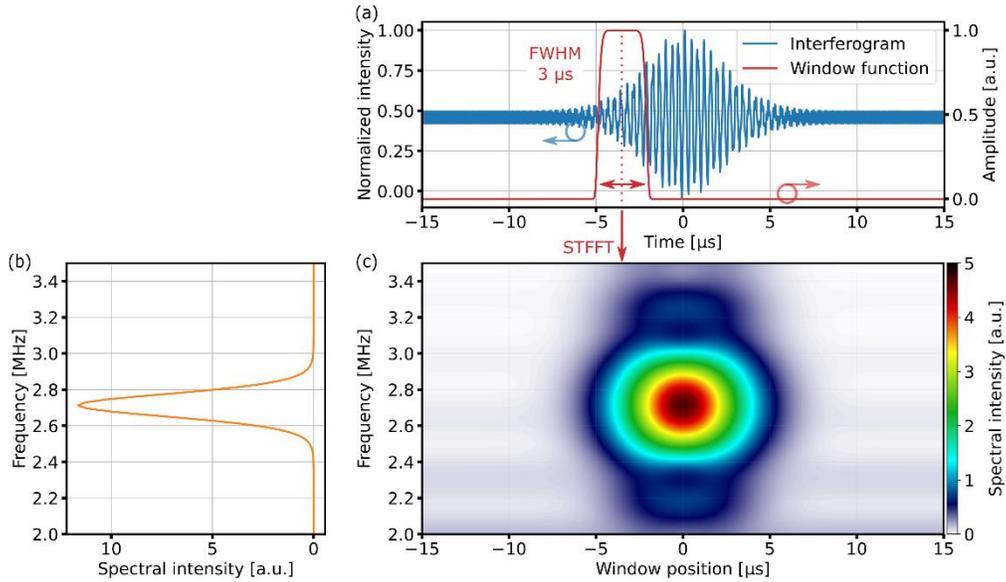

Fig. S1. (a) Simulated time-domain interferogram using two pulse trains with different repetition rates. (b) After FFT of the simulated time trace, we obtain a spectrum in the radio frequency domain in complete analogy to the measured trace. (c) Spectrogram generated by shifting the apodization time window across the interferogram and calculating the FFT, respectively.

In order to simulate the effect of chirp on our interferogram, we add the time-dependent frequency shift of self-phase modulation to the carrier frequency of our input pulses. This results in a broadened spectrum and the new frequency components lead to temporal wings in the spectrogram (see Fig. S2). We identify an elliptical-shaped pattern on top of the center part of the spectrogram, similar to the measured curve, as the wings converge to the center frequency $f_c^{RF}$ for larger $\tau$ due to the decreasing frequency shift of self-phase modulation (see Fig. S2c).

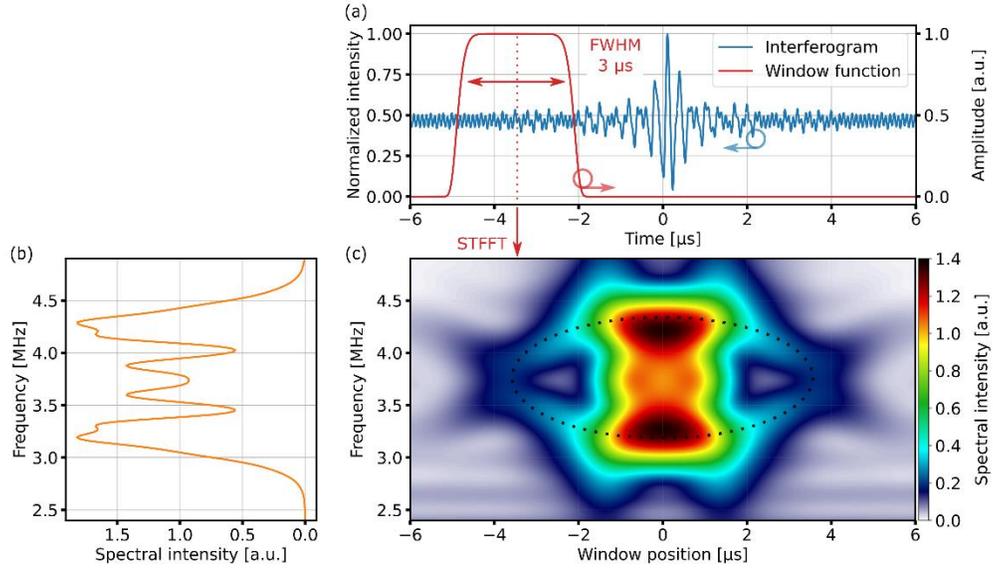

Fig. S2. (a) The simulated time-domain interferogram reveals less oscillations if we use chirped input pulses. (b) The FFT of the simulated time trace exposes a more complex structure than the previous case due to the new frequency components. (c) The spectrogram reveals a symmetric shape with respect to the center frequency. An elliptical structure is observed (indicated by white dashed line).

In addition, we perform the simulation with pulse trains that do not have equal amplitude values. This represents the configuration of our UV DCS system as we use different seed powers in front of each fiber. The target input pulse energy was 1.7 nJ (average power of 17 mW) with an output pulse energy of 0.43 nJ (4.3 mW) after the fiber.

In this case, the spectrogram is tilted and the spectral wings become asymmetric (see Fig. S3). This is in good agreement with our measured curve. Discrepancies in the shape and intensity distribution of the spectrogram arise because of inevitable deviations of the simulation parameters compared to the real values, especially fiber nonlinearity, dispersion and the input electric field of the UV laser pulse.

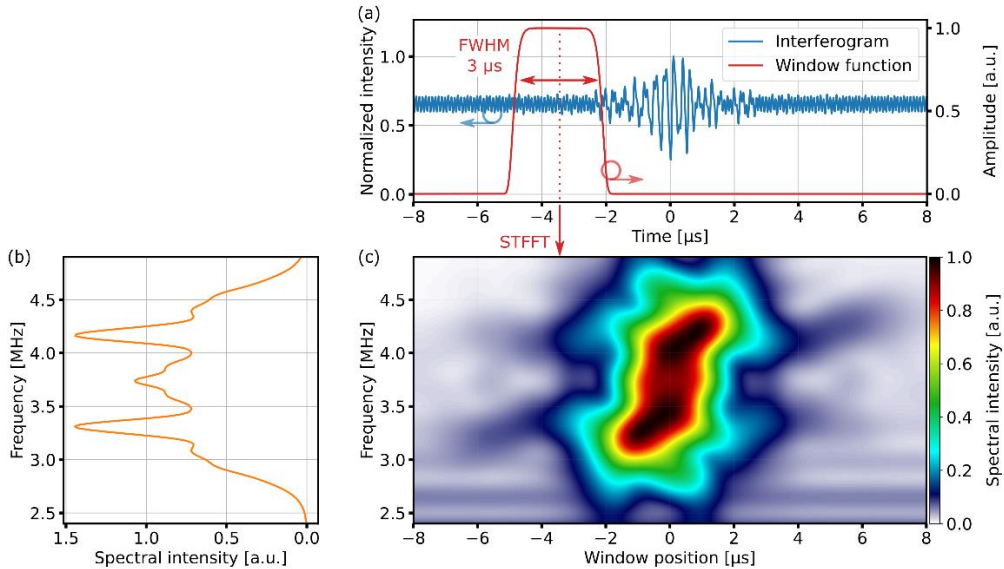

Fig. S3. (a) The simulated time-domain interferogram of unbalanced, chirped input pulses reveals more oscillations than the balanced case. The amplitude of the peak value decreases compared to the background signal. (b) The FFT of the simulated time trace shows a different intensity distribution than the balanced case. (c) The center part of the spectrogram is tilted and the spectral wings do not have equal spectral intensity.

**Formaldehyde preparation**

Figure S4 depicts the setup of the vacuum system attached to the gas sample cell (sample pathlength: 365 cm), which was used for the absorption experiments. The para-formaldehyde reservoir was heated up to 80-90°C and connected to the other components. We add a cold trap after the reservoir to reduce unwanted constituents with smaller condensation temperature than gaseous formaldehyde, e.g. polymer residuals. Piezo-based pressure gauges were employed to track the pressure. The repeatability of the pressure gauges is specified as 2 % of reading. No carrier gas was used during the preparation. The value of the pressure gauge (total pressure) yields the pressure of monomeric formaldehyde.

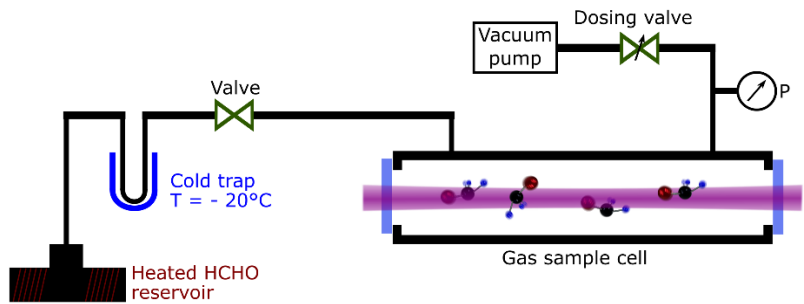

Fig. S4. Vacuum system for formaldehyde absorption experiments. The para-formaldehyde reservoir is connected to the sample cell via PTFE tubes. A vacuum pump in combination with (dosing) valves were employed to control the amount of formaldehyde in the sample cell. Pressure gauges (P) allow calculating the expected absorbance of the sample. The windows are UV-grade fused silica with a UV-VIS antireflection coating with flat spectral transmission, i.e. less than 1 % deviation.

**Dual-comb measurement duration and figures of merit**

The inverse of the detuning of our frequency combs defines the maximum apodization time window per interferogram for Fourier transformation, which also defines the maximum spectral resolution. For the dual-comb absorption experiments we set the detuning to 0.75 Hz for optimized spectrometer performance with regard to spectral resolution and to prevent aliasing effects. Following the noise analysis by Newbury et al., we are mainly limited by detector noise, which was estimated by the noise equivalent power ($NEP = 23.4 \times 10^{-11} \frac{W}{\sqrt{Hz}}$) of the photodetector [S1]. As the spectral information is mainly concentrated in a 4 µs interval (see Fig. 3d), using longer apodization time windows, i.e. higher spectral resolution, leads to a higher noise level. The apodization time window can be chosen shorter for optimized balance between signal-to-noise ratio and spectral resolution. In our case, we choose 100 µs for single-trace spectra and 270 µs for the averaged case. Averaging 6 single traces, using a 270 µs apodization time window, results in 8 s total acquisition time. This total acquisition time includes the total apodization time and waiting time [S2]. We analyzed the precision of the pressure determination using these spectra and we determined a single-shot value of 9.4 %.

The length of the time trace defines the resolution in the spectral domain. Here, these values are $\frac{1}{100 \mu s} \approx$ 10 kHz for the single trace and $\frac{1}{270 \mu s} \approx$ 3.7 kHz for the averaged case in the radio frequency domain. Multiplying with the upconversion factor $\frac{f_{rep,II}}{\delta} = \frac{10 \text{ MHz}}{0.75 \text{ Hz}}$ yields the resolution in the optical domain, which is 134 GHz (single trace) and 50 GHz (averaged) [S3]. The Doppler linewidth of formaldehyde at room temperature is on the order of a few GHz and is currently smaller than the resolution of our spectrometer [S4].

In order to quantify the SNR for the quality factor of the averaged absorbance curve, we employ the reference and sample spectrum that generate the absorbance spectrum in Figure 4a. We calculate the standard deviation of the transmission spectrum in the frequency region between 862.3 THz to 867.5 THz in complete analogy to Newbury et al. (see Figure 1c in Ref. 1). The inverse of this value yields the spectral signal-to-noise ratio and amounts to 23.3 for the averaged case. Experiments with an empty sample cell yield similar spectral SNR, which was calculated via the standard deviation over the full width of the spectrum. Together with the value of $M_T = 252 \sqrt{Hz}$ ($\Delta f$ = 35.7 THz, $\delta f$ = 50 GHz, T = 8 s), we achieve a quality factor of $5.9 \times 10^3 \sqrt{Hz}$. This value is about two orders of magnitude smaller than a theoretical calculation of the QF that has been previously performed by the authors (see Ref. [S5]). This deviation is mainly due to differences in the repetition frequencies and the noise equivalent power of the photodetector. In a further comparison, the work by Xu et al.

(see Ref. [47] in the main text) achieves a quality factor of $2.1 \times 10^3 \sqrt{Hz}$ ($\Delta f$ = 50 GHz, $\delta f$ = 500 MHz, T = 267 s, SNR = 345).

Calculating the quality factor of the single-trace measurement (see Fig. 4b) using the same procedure, yields a value of $3.9 \times 10^3 \sqrt{Hz}$ ($\Delta f$ = 35.7 THz, $\delta f$ = 133 GHz, T = 1.33 s, SNR = 16.8). The single-trace measurement features a shorter acquisition time window resulting in a lower SNR, but leading to similar quality factors for both cases. We observe a square-root-scaling of the SNR with respect to averaging time. We chose the averaging time such that the three vibronic branches are visible within the shortest acquisition time window possible. Further improvements of the averaging time and spectral resolution can be enabled by improving the frequency comb stability.

The noise level of our fast photodiode (see Figure 2) was calculated using the low-pass filter bandwidth of BW = 5 MHz in our setup (see Figure 1). This yields a power value of $P_{NEP} = NEP \cdot \sqrt{BW} = 23.4 \times 10^{-11} \frac{W}{\sqrt{Hz}} \cdot \sqrt{5\ MHz} \approx 0.52\ \mu W$. Converting this value to spectral intensity, using the spectral bin size $\delta_f = 81\ GHz$ in Figure 2b, gives the noise equivalent power of the fast photodiode $NEP_{FPD} = \frac{P_{NEP}}{\delta_f} \approx 6.5\ \mu W/THz$.

**Real-time monitoring of gas concentration changes**

The measured absorbance is in good agreement with the prepared concentration in the cell, and we observe a high degree of repeatability. The central frequency of our dual-comb spectrum is verified using a commercially-available grating spectrometer and literature data [S6]. In order to calculate the limit of detection (LOD), we determine the peak-absorbance-to-noise-ratio of our absorbance spectrum in the same way as we calculated the spectral signal-to-noise-ratio before (see Supplementary Note 3). We determine the LOD by division of the pressure used in this measurement by the peak-absorbance-to-noise-ratio, resulting in $p_{LOD}$=0.21 mbar.

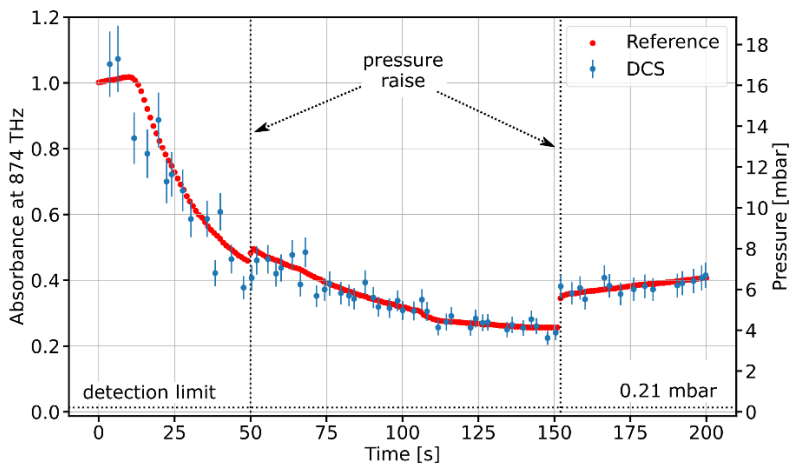

Fig. S5. Time-resolved absorbance values of formaldehyde at 874 THz measured with DCS (blue dots) and with a pressure gauge (reference, red dots). The uncertainty bars indicate a single-shot precision of 9.4 %, and we assume constant uncertainty within the specified measurement range. We intentionally introduce concentration changes after 50 s and 152 s. The absorbance values retrieved from the UV DCS measurements coincide with the reference points with a root-mean-square deviation of 7.9 % proving its applicability for gas monitoring. The single-trace detection limit is 0.21 mbar (horizontal dotted line).

For instant time tracking DCS measurements, we fill the sample cell with gaseous formaldehyde to a pressure of 16 mbar, see Fig. S5). Then we close the valve to the reservoir and evacuate the cell, while acquiring interferogram traces. We then calculate the absorbance spectra using the reference and sample spectrum. To facilitate comparison between subsequent measurements, we subtract a baseline for each absorbance spectrum in order to ensure a baseline at zero absorbance. The smallest pressure in the cell during this time series is 4.1 mbar, at about t = 140 s. We chose the peak value at 874 THz to calculate the absorbance of the dual-comb measurement and the expected absorption via the pressure in the cell (see Fig. S5). After an arbitrarily chosen time, in this scan after 50 s and 152 s, the valve to the reservoir was opened to intentionally introduce changes in the sample concentration, which was probed by dual-comb spectroscopy (indicated by vertical dotted lines in Fig. S5). The DCS and pressure gauge measurements are in excellent agreement (root-mean-square deviation of 7.9 %) and prove the potential of UV DCS for future investigations of complex reaction dynamics. The single-shot precision demonstrated here (see Fig. S5) is within a factor of two compared to commercially-available devices certified for work safety control qualifying for typical environmental monitoring needs. Averaging multiple traces and stabilization of our setup will improve the precision further. The high absorption cross section of formaldehyde in the UV yields a single-trace detection limit of 0.21 mbar.

We calculate the noise equivalent absorption NEA using the absorption coefficient α of $5.1 \times 10^{-4}\ cm^{-1}$ corresponding to 0.21 mbar formaldehyde via an absorption cross section of $1.0 \times 10^{-19}\ cm^2$/molecule. This yields a noise equivalent absorption of $NEA = \alpha \cdot \sqrt{T} = 5.1 \times 10^{-4}\ cm^{-1} \cdot \sqrt{1.33\ s} = 5.9 \times 10^{-4}\ cm^{-1}\ Hz^{-1}$. The noise equivalent absorption per spectral channel amounts to $NEA_M = \frac{NEA}{\sqrt{M}} = \frac{5.9 \times 10^{-4}\ cm^{-1}\ Hz^{-1}}{\sqrt{\frac{35.7\ THz}{133\ GHz}}} = 3.6 \times 10^{-5}\ cm^{-1}\ Hz^{-1}$, where M is the number of resolution elements.

The sensitivity achieved here qualifies our system for sensing industrial and environmental applications. Other techniques, e.g. broadband cavity-enhanced absorption spectroscopy in the ultraviolet [S7], achieve higher sensitivity. Cavity enhanced DCS has been implemented [S8] and has also the potential to improve the sensitivity of UV DCS [S9].

**Analysis of dual-comb absorption spectra**

We compare the spectra obtained directly from the interferogram traces of a single reference and sample scan. The reference spectrum ($\tilde{V}_{ref}$) is multiplied by a constant factor to account for the difference in signal strength between the reference and the sample path. Notably, there is a one-to-one reference possible for each transmission spectrum as the reference and sample interferogram are generated by identical pulses. Before the experiment, it was verified that the spectra from the reference path and from the sample path plus empty cell do not exhibit discrepancies in intensity across the full spectrum. Then, we determine the absorbance A using the equation $A = -\log_{10} T$ with the transmission $T = \frac{\tilde{V}}{\tilde{V}_{ref}}$, where $\tilde{V}$ is the sample spectrum obtained after FFT of the sample path interferogram. The absorbance spectrum of these two curves directly features a flat baseline, e.g. in the frequency range from 853 THz to 865 THz. Experiments with an empty cell also result in a flat baseline in the absorbance spectrum with equal noise level as compared to the measurement with a sample. This proved that no further baseline corrections are required. The reference spectrum was multiplied by a factor of 2.2 not changing the SNR or the baseline noise (see Fig. S6). Experiments with an empty sample cell confirmed that the baseline is flat over the entire spectral range. The FFT of both spectra can be obtained in less than 200 ms and we exclude spectra that do not reach beyond the upper formaldehyde rovibronic branch (f > 891 THz).

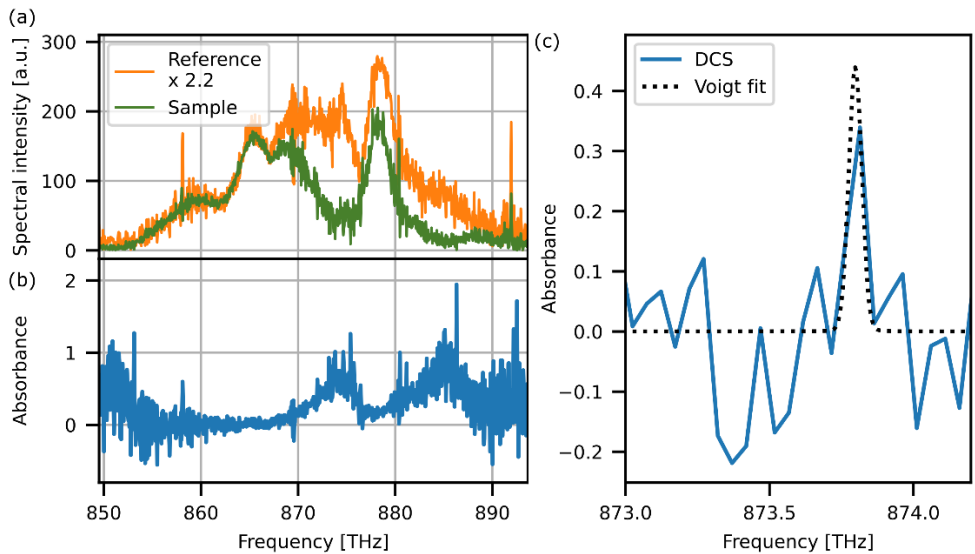

Fig. S6. (a) Single-trace reference (orange) and sample (green) spectrum in the ultraviolet spectral region measured with dual-comb spectroscopy at 50 GHz spectral resolution. Both spectra were obtained by computing the FFT of the reference and sample interferogram, respectively. The shape of both spectra is very similar such that the baseline can be determined directly after calculating the absorbance. (b) Resulting single-trace absorbance spectrum with 50 GHz resolution of the two traces in (a). (c) Line fit of one rovibronic transition using a Voigt line shape.

We identify the transitions in the absorbance spectrum in Fig. 4a. The uncertainties were calculated using the six traces with 270 μs apodization time window length. Performing a line fit confirms the spectral resolution of 50 GHz (see Fig. S6c). We use a Voigt line shape, but we expect the instrumental response function to dominate the line shape. The quantum numbers with the corresponding frequencies are listed in table S1 [S10].

Table S1. Six prominent transitions identified with UV DCS

| $\widetilde{A}^1A_2$-$\widetilde{X}^1A_2$ electronic transition | | | |
|---|---|---|---|
| Vibronic transition | K´´ | J´´ | Frequency [THz] |
| $4^3_0$ | 3 | 15 | $873.8 \pm 0.1$ |
| $4^3_0$ | 3 | 8 | $874.6 \pm 0.1$ |
| $4^3_0$ | 3 | 5 | $875.4 \pm 0.1$ |
| $2^1_0 4^1_0$ | 4 | 9 | $884.5 \pm 0.2$ |
| $2^1_0 4^1_0$ | 0 | 11 | $885.3 \pm 0.2$ |
| $2^1_0 4^1_0$ | 0 | 6 | $886.2 \pm 0.1$ |